\documentclass[sigconf]{acmart}
\usepackage{subcaption}

\setcopyright{none}
\AtBeginDocument{%
  \providecommand\BibTeX{{%
    \normalfont B\kern-0.5em{\scshape i\kern-0.25em b}\kern-0.8em\TeX}}}

\copyrightyear{}
\acmDOI{}

\acmConference{}
\acmBooktitle{}
\acmPrice{}
\acmISBN{}

\acmSubmissionID{}

\begin{document}

\title[Mapping New Informal Settlements using Machine Learning and Time Series Satellite Images]{Mapping New Informal Settlements using Machine Learning and Time Series Satellite Images: An Application in the Venezuelan Migration Crisis}

\author{Isabelle Tingzon}
\email{issa@thinkingmachin.es}
\affiliation{%
  \institution{Thinking Machines Data Science}
  \city{Taguig City}
  \country{Philippines}
}

\author{Niccolo Dejito}
\email{cholo@thinkingmachin.es}
\affiliation{%
  \institution{Thinking Machines Data Science}
  \city{Taguig City}
  \country{Philippines}
}

\author{Ren Avell Flores}
\email{avell@thinkingmachin.es}
\affiliation{%
  \institution{Thinking Machines Data Science}
  \city{Taguig City}
  \country{Philippines}
}

\author{Rodolfo De Guzman}
\email{ram@thinkingmachin.es}
\affiliation{%
  \institution{Thinking Machines Data Science}
  \city{Taguig City}
  \country{Philippines}
}

\author{Liliana Carvajal}
\email{lcarvajal@immap.org}
\affiliation{%
  \institution{IMMAP Colombia}
  \city{Bogota}
  \country{Colombia}
}

\author{Katerine Zapata Erazo}
\email{kerazo@immap.org}
\affiliation{%
  \institution{IMMAP Colombia}
  \city{Bogota}
  \country{Colombia}
}

\author{Ivan Enrique Contreras Cala}
\email{icontreras@immap.org}
\affiliation{%
  \institution{IMMAP Colombia}
  \city{Bogota}
  \country{Colombia}
}

\author{Jeffrey Villaveces}
\email{jvillaveces@immap.org}
\affiliation{%
  \institution{IMMAP Colombia}
  \city{Bogota}
  \country{Colombia}
}

\author{Daniela Rubio}
\email{daniela@premise.com}
\affiliation{
  \institution{Premise Data}
  \city{San Francisco}
  \state{California}
}

\author{Rayid Ghani}
\email{rayid@cmu.edu}
\affiliation{%
  \institution{Carnegie Mellon University}
  \city{Pittsburgh}
  \state{Pennsylvania}
  }

\renewcommand{\shortauthors}{Tingzon, et al.}

\begin{abstract}
  Since 2014, nearly 2 million Venezuelans have fled to Colombia to escape an economically devastated country during what is one of the largest humanitarian crises in modern history. Non-government organizations and local government units are faced with the challenge of identifying, assessing, and monitoring rapidly growing migrant communities in order to provide urgent humanitarian aid. However, with many of these displaced populations living in informal settlements areas across the country, locating migrant settlements across large territories can be a major challenge. To address this problem, we propose a novel approach for rapidly and cost-effectively locating new and emerging informal settlements using machine learning and publicly accessible Sentinel-2 time series satellite imagery. We demonstrate the effectiveness of the approach in identifying potential Venezuelan migrant settlements in Colombia that have emerged between 2015 to 2020. Finally, we emphasize the importance of post-classification verification and  present a two-step validation approach consisting of (1) remote validation using Google Earth and (2) on-the-ground validation through the Premise App, a mobile crowdsourcing platform.
\end{abstract}





\maketitle

\section{Introduction}

Since the economic and sociopolitical downturn of Venezuela in 2014, more than 4.5 million Venezuelans are estimated to have fled the country, resulting in one of the largest forced displacements in Latin America's recent history \cite{worldbank2019,freier2018understanding}. Facing spiraling hyperinflation and scarcity of basic necessities, more than 1.8 million migrants Venezuelans have crossed onto the bordering country Colombia; unfortunately, many of these migrants still struggle to survive as they face poverty, unemployment, food insecurity, and health problems, exacerbated only further by the ongoing COVID-19 pandemic \cite{daniels2020venezuelan}. With thousands emigrating each day, non-government organizations (NGOs) and local government units (LGUs) are tasked with the responsibility of identifying, assessing, and monitoring rapidly growing migrant populations in order to provide urgent humanitarian assistance.

To obtain accurate information on migrant populations, humanitarian groups typically conduct field surveys and interviews \cite{immap2019}. Based on on-the-ground reports, many Venezuelan migrants live in informal settlements that share common characteristics such as small roof sizes; substandard housing material (e.g. plastic, cardboard, metal, and  natural materials); a disorganized and unstructured layout; and a lack of a nearby structured road network, signifying the absence of proper urban planning. Moreover, many of these settlements are located on the outskirts of more formal cities or towns, where there is closer proximity to potential employment and services. 

\begin{figure}[t!]
\begin{subfigure}{\linewidth}
\centerline{
\includegraphics[width=0.48\linewidth]{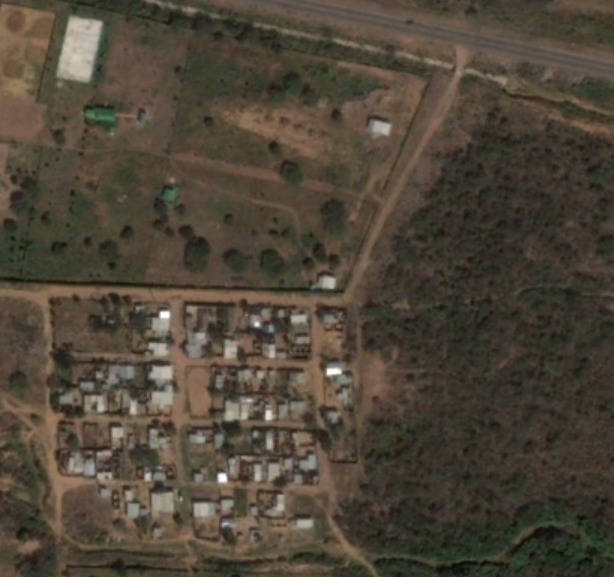}\hspace*{1mm}
\includegraphics[width=0.48\linewidth]{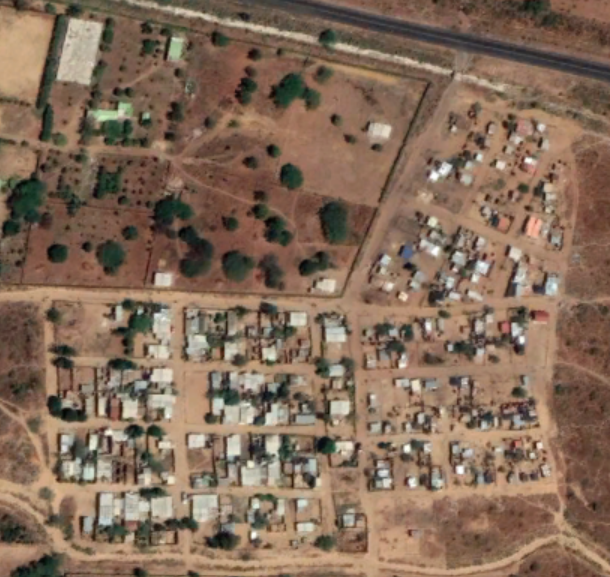}
}
\vspace*{2.5mm}
\end{subfigure} 
\begin{subfigure}{\linewidth}
\centerline{
\includegraphics[width=0.48\linewidth]{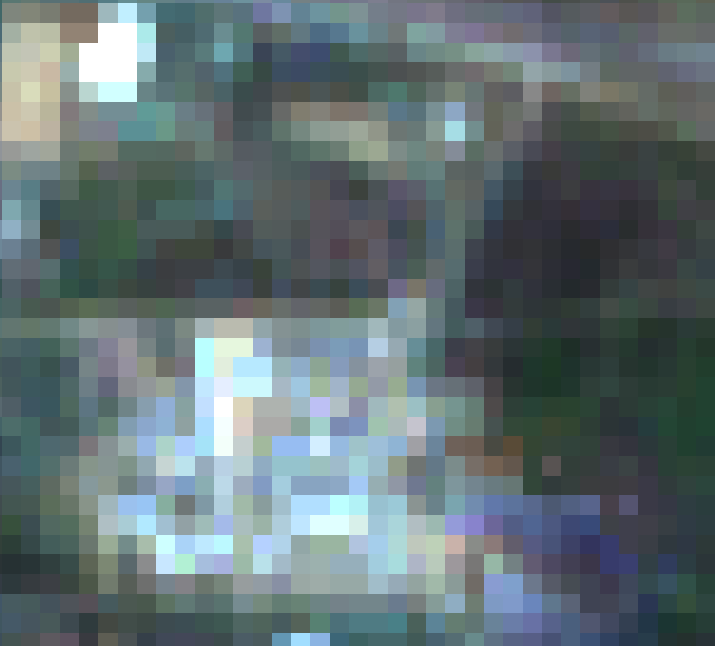}\hspace*{1mm}
\includegraphics[width=0.48\linewidth]{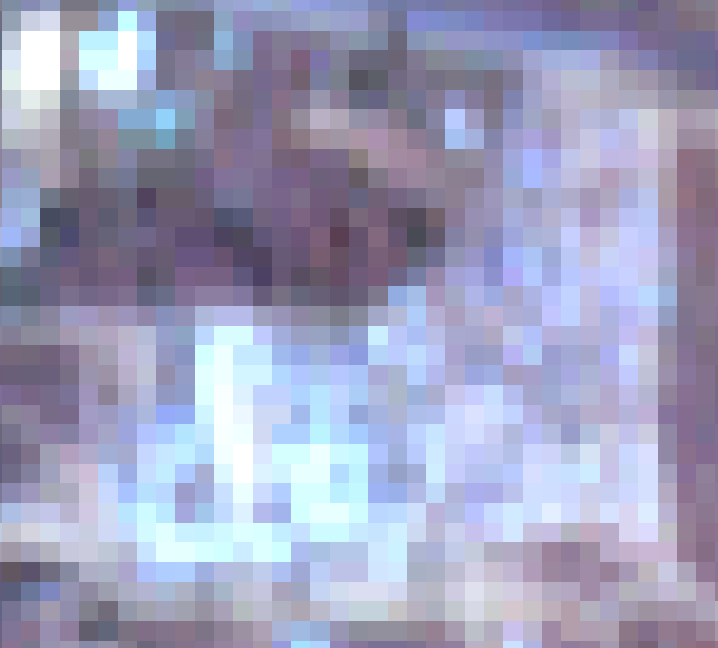}
}
\end{subfigure}
\caption{An example of two high-resolution Google Earth images (top) and two low-resolution Sentinel-2A images (bottom) of the same area, demonstrating the emergence of a new informal Venezuelan migrant settlement in Maicao, La Guajira from 2015 (left) to 2020 (right).}
\label{fig1:example}
\end{figure}

With many of these informal migrant settlements scattered across Colombia, conducting high coverage surveys can be challenging, especially given the limited manpower and resources available to non-profit organizations. In recent years, several research works have sought to use a combination of computer vision and satellite images to quickly and efficiently identify informal settlements and other vulnerable communities \cite{gram2019mapping, klemmer2020population, jean2016combining, mboga2017detection}. As shown in Figure \ref{fig1:example}, the emergence of informal settlements can indeed be visually observed in high resolution historical satellite images. However, the high costs associated with acquiring high resolution images coupled with compute-intensive deep learning approaches can be a barrier to adoption for many NGOs and LGUs. 

To address these challenges, we present an approach for rapidly and inexpensively locating new and emerging informal settlements, with the goal of helping humanitarian organizations focus their efforts in areas with higher likelihoods of housing migrant populations. More specifically, we propose to use low-resolution, high-temporal Sentinel-2A satellite images and non-computationally intensive machine learning methods to detect and map informal settlements in Colombia that have emerged after 2014, marking the inception of the Venezuelan mass migration crisis \cite{wilsoncenter2019}.

These methods are then integrated into an end-to-end pipeline that converts raw Sentinel-2A images to 10 m resolution probability maps that encode higher probability areas as brighter pixels. Using the informal settlement probability maps, we generate polygons which are then deployed onto a real-world crowdsourcing platform powered by Premise Data \cite{premise2019} in order to mobilize a network of on-the-ground validators to collect survey data in high-probability areas. 

To summarize, our main contributions are as follows:
\begin{itemize}
    \item We present a novel approach to mapping new and emerging informal settlements using a combination of cost-efficient machine learning-based methods and publicly accessible Sentinel-2A time series satellite images.
    \item We demonstrate the effectiveness of the approach in identifying potential informal Venezuelan migrant settlements in Colombia that have emerged between 2015 to 2020.
    \item We propose a two-step post-classification validation approach that involves (1) remote validation using GIS applications followed by (2) on-the-ground validation through the Premise App, a mobile crowdsourcing platform.
    \item We make the source code publicly available \cite{github2020}.
\end{itemize}

While the proposed approach is designed to help humanitarian organizations rapidly and cost-effectively locate vulnerable migrant communities, we emphasize the importance of post-classification on-the-ground validation. This is critical for decision making that involves the formulation of strategies and policies that significantly impact people's quality of life. 

\section{Related Work}
Recent years have seen an increase in the number of research works focusing on the applications of machine learning to remote sensing data for mapping vulnerable populations \cite{gram2019mapping, klemmer2020population, jean2016combining, perez2017poverty, xie2016transfer, head2017can, wu2020estimation, babenko2017poverty, wurm2017exploitation, maiya2018slum, mboga2017detection}. Many of these studies rely on the combination of computer vision and high resolution satellite images; for example, the  work by Jean et al. used a combination of high resolution daytime satellite images, nighttime luminosity data, and convolutional neural networks (CNNs) to produce granular poverty maps for sub-Saharan African countries \cite{jean2016combining}. Another study by Mboga and Persello used CNNs to automatically detect informal settlements in QuickBird images of Dar es Salaam, Tanzania \cite{mboga2017detection}. 

Given the high costs associated with the acquisition of high resolution satellite images as well as the development of computationally intensive deep learning models, several works have sought to use low resolution satellite images and traditional machine learning methods as an alternative, more cost-effective approach to poverty mapping \cite{perez2017poverty, gram2019mapping, wurm2017exploitation}. For example, the work by Gram-Hansen et al. demonstrates the viability of using freely available low-resolution Sentinel-2A data to detect informal settlements across different countries \cite{gram2019mapping}. Specifically, the study compares the performance of two approaches: (1) computationally efficient canonical correlation forests (CCFs) to learn the spectral signature of informal settlements in low-resolution satellite imagery and (2) CNNs to extract finer grained features in very high resolution satellite images. Another study by Wurm et al. extracted image texture features and morphological profiles to map slums in Mumbai, India using Sentinel-2A satellite images \cite{wurm2017exploitation}. 

Apart from being free and publicly accessible, low resolution remote sensing data also have a long time series which allows users to easily harness the advantages of historical satellite imagery. However, low resolution time series satellite imagery are often used in the context of crop monitoring and land cover change detection \cite{feng2019crop, rapinel2019evaluation, jiang2019early, kleynhans2012land} and have rarely been explored for its potential in identifying vulnerable communities in contexts where temporality plays a major role. In this study, we leverage low-resolution time series satellite images for the unique use case of identifying informal settlements that have emerged recently, i.e. between 2015 to 2020, as a means of mapping potential Venezuelan migrant communities across Colombia. To our knowledge, there exists no previous works exploring the use of low resolution time series satellite images for informal settlement detection. 

\section{Informal Settlement Definitions}
%For the purposes of the project, a series of criteria were established that allow a more precise definition for the new settlements, the first of which responds to a geographical criterion where the settlements are those that are located in great majority in the peripheral areas of the municipality, adjoining within the limits of the peri-urban zones, place of delimitation between the urban and the rural. It is also possible that a few settlements are located in rural territory, plus most of the settlements due to the dynamics of the conurbation will be found in urban areas.

In this section we describe a series of criteria to provide a more precise definition of informal migrant settlements. 

\subsubsection{Geography}
In term of geography, informal settlements are located mostly in the peripheral areas of the municipality, near the borders of peri-urban zones, i.e. places of delimitation between the urban and the rural. While many informal settlements are located in urban areas due to the dynamics of conurbation, some informal settlements may also be found in rural territory.

%Regarding the criterion of temporality, the settlements were selected from the period of analysis that between 2015 and 2020, this because the migratory phenomenon according to sources of Migration Colombia began to be more sensitive for the country since 2015, being much greater migratory flow in later years. Additionally, this fact is complemented by the fact that, at the methodological level, the validation process allows us to identify that these settlements are recent (less than three years of built and constituted in the territory).

\subsubsection{Temporality} We define migrant settlements as settlements that have emerged between 2015 and 2020, based on the migratory phenomenon in Venezuela that began in 2014 and progressed with a much greater migratory flow in later years \cite{colmig2017}. This is complemented by the fact that at the methodological level, the validation process, described in later sections, allows us to more easily determine whether or not these settlements are recent, i.e. built less than three years ago and constituted in the territory.

%The third criterion responds to social dynamics, where new settlements are made up of one or several groups of individuals who have established themselves in undeveloped areas and close to the urban layout, which is why their treatment often maintains characteristics of informal settlements and with conditions of vulnerability and poverty. These groups are possibly made up in part of the Venezuelan migrant and refugee population, taking into account the migratory dynamics, information from official sources and previous studies that have characterized this situation.

\subsubsection{Social Dynamics} In terms of social dynamics, we define new informal settlements as settlements that are made up of one or more groups of individuals that have established themselves in underdeveloped areas near urban regions in conditions of vulnerability and poverty. These groups are in part made up of the Venezuelan migrant and refugee population \cite{colmig2017}.

%Regarding the physical criteria for the new settlements, the vulnerability conditions of informal and precarious settlements were considered, such as:
%\begin{itemize}
%    \item The absence of urban furniture (Parks, communal rooms, public lighting, platforms, etc).
%    \item Messy layout, nonexistence of a (paved) road network and irregular division of the properties.
%    \item Improvised houses (made with possible perishable materials such as wood, brass, etc.).
%\end{itemize}
%These physical characteristics correspond to that related to Document Conpes 3604 .

\subsubsection{Physical Characteristics} Consistent with the physical characteristics described in the Document Conpes 3604 \cite{conpes3604}, we consider the the vulnerability conditions of new informal settlements such as:
\begin{itemize}
    \item The absence of urban indicators in the environment such as parks, communal rooms, public lighting, platforms, etc.
    \item Disorganized layout, nonexistence of a paved road network, and irregular division of the properties.
    \item Improvised houses, made with perishable materials such as wood, brass, etc..
\end{itemize}

%Finally, the new settlements are not: shelters, places of passage, points of attention, camps, refuges, and in general all infrastructure that has been formed by some governmental, non-governmental or supranational entity; that is to say, that they must be the initiative of the same community or population and that their temporally is not planned in the short term.

Finally, the new settlements are not: shelters, places of passage, points of attention, camps, refuges, and in general all infrastructure that has been formed by some governmental, non-governmental or supranational entity; that is to say, that they must be the initiative of the same community or population and that their temporally is not planned in the short term.

\section{Data}
\subsection{Sentinel-2 Data}
Launched in 2015, the Sentinel-2A satellite is among the first wide-swath, multispectral imaging missions under the Copernicus program by the European Commission (EC) in partnership with the European Space Agency (ESA). With its global coverage and frequent revisit time, Sentinel-2A is used in a wide array of applications including land cover monitoring, agriculture and forestry, humanitarian relief operations, disaster risk mitigation, and security \cite{sentinel2user}. Sentinel-2A data is free and publicly available with moderate spatial resolution and a long time series, making it easy for users to leverage the advantages of historical satellite imagery and explore new opportunities in earth observation.

Sentinel-2A images contain 13 spectral bands with varying spatial resolutions ranging from 10 m to 60 m. Specifically, four bands have 10 m resolution: blue (b2: 490 nm), green (b3: 560 nm), red (b4: 665 nm) and near-infrared (b8: 842 nm); six bands have 20 m resolution: red edge (b5: 705 nm, b6: 740 nm, b7: 783 nm, b8A: 865 nm) and shortwave infrared (b11: 1610 nm, b12: 2190 nm); and three atmospheric bands have 60 m resolution (b1: 443 nm, b9: 940 nm, b10: 1375 nm). 

Two types of Sentinel-2 images are available to users: Level-1C and Level-2A. Level-1C products are radiometrically scaled, orthographically projected, and geometrically corrected, with top-of-atmosphere reflectances. Level-2A products are similar to Level-1C with the exception of being atmospherically corrected resulting in bottom-of-atmosphere reflectance. 

In this work, we acquired Sentinel-2A data via Google Earth Engine (GEE) and generated a time series of biennial composites of satellite images from 2015 to 2020 for each of the nine municipalities. More specifically, we generated a single median-aggregated composite for each of the following year ranges: 2015 to 2016, 2017 to 2018, and 2019 to 2020. Biennial composites were preferred over annual composites as they contained less cloud cover per image. To generate the composites, we calculated the median of all cloudless pixels available within the two years and resampled each band to the highest geometric resolution of 10 m. Note that due to the limited availability of Level-2A products in GEE, we used Level-1C images for years 2015 to 2017 and Level-2A images for years 2018 to 2020. 

\subsection{Ground Truth Annotations}
\label{ref:ground-truth}
In this work, we used field data of informal migrant settlements collected across the different regions of Colombia by the humanitarian organization iMMAP in 2019. The dataset contains a total of 36 ground-validated coordinate locations of informal migrant settlements within the nine municipalities: Maicao, Riohacha, Uribia, Arauca, Arauqita, Tibu, Cucuta, Soacha, and Bogota. For each coordinate, we generated a vector polygon enclosing the informal settlement area. We then projected these polygons onto the Sentinel-2A images to obtain 10 m resolution raster masks and extracted historical spectral information at these specific pixels. The number of positive pixels per municipality are shown in Table \ref{table:distribution}.

\begin{table}[h]
\caption{Number of positive pixels and polygons per municipality}
\begin{center}
\begin{tabular}{lcc}
\hline
\multicolumn{1}{l}{\textbf{Municipality}} & \multicolumn{1}{c}{\textbf{\begin{tabular}[c]{@{}c@{}}Positive\\ Pixels\end{tabular}}} &
\multicolumn{1}{c}{\textbf{\begin{tabular}[c]{@{}c@{}}Positive\\ Polygons\end{tabular}}} \\
\hline
Arauca & 2,298 & 7 \\ 
Arauquita & 778 & 2 \\ 
Bogota & 2,720 & 6 \\ 
Cucuta & 2,485 & 3 \\
Maicao & 552 & 7 \\ 
Riohacha    &   3,501   & 4 \\ 
Soacha & 347 & 1\\
Tibu & 730 & 3\\
Uribia      &   10,345      &     3  \\
\hline
\end{tabular}
\label{table:distribution}
\end{center}
\end{table}

\begin{figure}[h]
\begin{subfigure}{0.4\columnwidth}
\centerline{
\includegraphics[width=\linewidth]{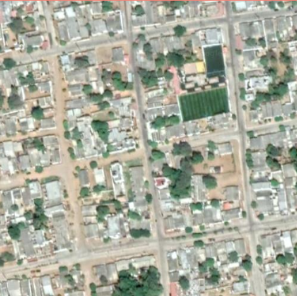}
}
\caption{Formal settlement}
\end{subfigure} \hspace*{1mm}
\begin{subfigure}{0.4\columnwidth}
\centerline{
\includegraphics[width=\linewidth]{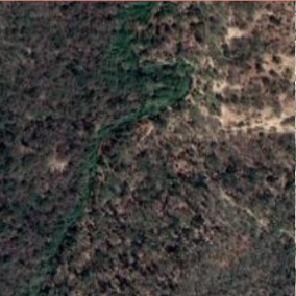}
}
\caption{Unoccupied land area}
\end{subfigure}
\par\bigskip
\begin{subfigure}{0.8\columnwidth}
\centerline{
\includegraphics[width=\linewidth]{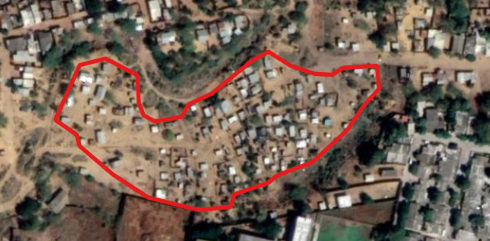}
}
\caption{Informal settlement}
\end{subfigure}
\caption{Examples of formal settlement grid (top left), an unoccupied land area (top right) and an informal settlement enclosed in a polygon (bottom)}
\label{fig1:classes}
\end{figure}

To form our set of negative examples, we looked towards random sampling: for each municipality, we generated 500 m x 500 m grid blocks and randomly selected a minimum of 30 grids. To ensure that a sufficient number of formal settlement examples are represented in the dataset, we manually sampled at least three grids within the urban city center of each municipality.  Through visual interpretation of high-resolution historical images available on Google Earth, we examined each selected grid to ensure it contained only formal settlements or unoccupied land masses as illustrated in Figure \ref{fig1:classes}, i.e. the grid must not contain any visual characteristics associated with informal settlement areas such as small, disorganized rooftops; otherwise we remove the grid and select an alternative that satisfies this criteria. 

\section{Methods}

\subsection{Sampling Strategy}
We sampled a total of 293,756 pixels, comprised of 23,756 positive samples and 270,000 negative samples. Due to the sparsity of the ground truth informal settlements data, we included all positive pixel samples available for each of the nine municipalities. For the negative examples, we sampled a total of 30,000 negative pixels per municipality, split between formal settlements (40$\%$) and unoccupied land areas (60$\%$). 

\subsection{Feature Extraction}
From the 13 spectral bands, we selected bands b1, b2, b3, b4, b5, b6, b7, b8, b8A, b9, b11, and b12 as input features. The cirrus band b10 is excluded from Level-2A products as it does not contain surface information \cite{sentinel2user}. 

For each annual composite, we also derived several vegetation and built-up indices commonly used in land cover classification and urban growth analysis \cite{valdiviezo2018built}. One such index is the normalized difference vegetation index (NDVI), which quantifies vegetation cover by calculating the difference between the red and near-infrared (NIR) bands \cite{rouse1974monitoring}. NDVI values lie within the range $[-1, 1]$ with values closer to 1 indicating higher density of green leaves. The soil adjusted vegetation index (SAVI) is used to detect vegetation in low plant density areas and introduces the adjustment factor $\ell$, which ranges from 0 (high vegetation cover) to 1 (low vegetation cover) \cite{huete1988soil}. The modified normalized difference water index (MNDWI) uses green and shortwave infrared (SWIR) bands to detect open water features \cite{xu2006modification}.

A number of different built-up indices have also been proposed in recent literature \cite{valdiviezo2018built}. The most common is the normalized difference building index (NDBI), which highlights urban areas with higher reflectance response in the SWIR band compared to the NIR band \cite{zha2003use}. The urban index (UI) exploits the inverse relationship between the NIR and mid-infrared portions of the spectrum in order to highlight urban spread \cite{kawamura1996relation}. The new built-up index (NBI) is computed using the spectral response of different land covers in the red, NIR, and SWIR bands \cite{jieli2010extract}. The band ratio for built-up areas (BRBA) and new built-up area index (NBAI) were both introduced by Waqar et al. to extract bare soil and built-up areas using medium resolution Landsat imagery \cite{waqar2012development}. Similarly, the modified built-up index (MBI) was proposed to improve the delineation between built-up areas and other land as a means to quantify urban sprawl \cite{liu2014spatiotemporal}. Finally, the built-up area extraction index (BAEI) was introduced as a new spectral index that uses red, green, and SWIR bands to extract built-up areas in Landsat-8 images \cite{bouzekri2015new}.  

We summarize the calculation formula of the derived indices in Table \ref{table:index}. 

\begin{table}[htbp]
\caption{Derived Index Calculation Formula}
\begin{center}
\begin{tabular}{cc}
\hline
\multicolumn{1}{c}{\textbf{Derived Index}} & \multicolumn{1}{c}{\textbf{Formula}} \\ \hline
\rule{0pt}{2.5ex}   
NDVI                                       &
$\frac{b_8 - b_4}{b_8 + b_4}$                    
\\
\rule{0pt}{2.5ex}
SAVI                                       &  
$\frac{(b_{8A}-b_4) (1 + \ell) }{b_{8A} + b_4 + \ell}$
\\
\rule{0pt}{2.5ex}
MNDWI                                       &  
$\frac{b_3-b_{11}}{b_3+b_{11}}$
\\
\rule{0pt}{2.5ex}  
NDBI                                       & 
$\frac{b_{11}-b_{8}}{b_{11}+b_{8}}$
\\ 
\rule{0pt}{2.5ex} 
UI                                      &  
$\frac{b_7-b_5}{b_7+b_5}$
\\ 
\rule{0pt}{2.5ex} 
NBI                                      &  
$\frac{b_4 \cdot b_{11}}{b_{8a}}$
\\ 
\rule{0pt}{2.5ex} 
BRBA                                    &  
$\frac{b_4}{b_{11}}$
\\ 
\rule{0pt}{2.5ex} 
NBAI                              &  
$\frac{b_{11}-b_{12}/b_3}{b_{11}+b_{12}/b_3}$
\\ 
\rule{0pt}{2.5ex} 
MBI                           &  
$\frac{b_{12} \cdot b_4-b_{8A}^2}{b_4+b_{8A}+b_{12}}$
\\ 
\rule{0pt}{2.5ex} 
BAEI                           &  
$\frac{b_4+c}{b_3+b_{11}}$
\\
\hline
\end{tabular}
\label{table:index}
\end{center}
\end{table}

\subsection{Model Training and Evaluation}
 We modeled the problem as a supervised pixel-wise classification task and extracted a total of 66 features per pixel: 12 raw spectral bands and 10 derived indices for each of the three biennial composites from 2015 to 2020. We then compared the performance of the following machine learning models: logistic regression, linear support vector machines (SVM), and random forest. For model evaluation, we used a spatial cross validation (CV) approach in the form of leave-one-municipality-out CV to overcome data leakages brought about by spatial autocorrelation and to measure the model's generalizability when applied to new and unseen geographies. For model training, we leveraged a Google Compute Engine (GCE) instance with 4 vCPUs and 15 GB of memory (n1-standard-4).
 
\begin{figure*}[h]
  \centering
  \includegraphics[width=\linewidth]{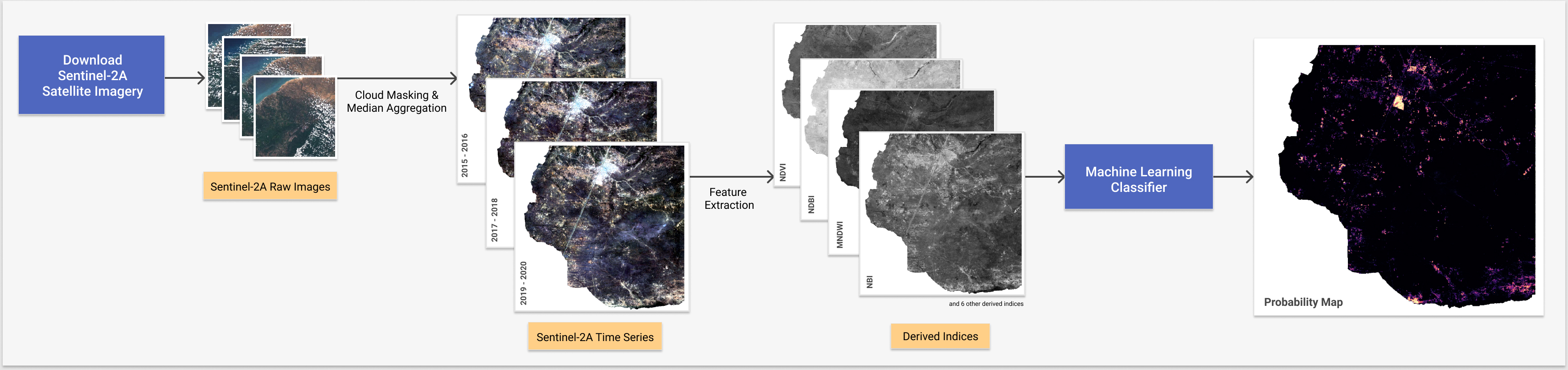}
\caption{An end-to-end pipeline that converts low resolution Sentinel-2A time series data for any region of interest (ROI) into an informal settlement probability map.}
\label{fig:pipeline}
\end{figure*}

 The output of the trained model, is a probability map that encodes higher probability areas as brighter pixels. Figure \ref{fig:pipeline} illustrates the end-to-end pipeline for converting time series Sentinel-2A imagery to pixel probability maps. These probability maps are then used to efficiently guide the identification of informal settlements.
 
\section{Results and Discussion}

\subsection{Performance Assessment}
To evaluate model performance, we compute both precision and recall. Precision measures the proportion of positively identified samples that is correct; recall measures the proportion of actual positive samples that is correctly identified by the model. We compute precision and recall using standard definitions as follows:
\begin{equation}
    \text{Precision} = \frac{tp}{tp + fp}
\end{equation}
\begin{equation}
    \text{Recall} = \frac{tp}{tp + fn}
\end{equation}
where $tp$ is the number of true positives, $fp$ the number of false positives, and $fn$ the number of false negatives.

We compute both pixel-level and settlement-level precision and recall for each out-of-sample municipality in the dataset. We define a "settlement" as a group of pixels whose pixel probabilities are aggregated to form a settlement-level probability. Specifically, we group pixels based on their membership to either an informal settlement polygon (positive settlement) or a negative 500 m x 500 m grid block (negative settlement), the construction of which is described in Section \ref{ref:ground-truth}. Aggregation was done by computing the mean of the top 10$\%$ pixel probabilities per settlement. The intuition behind computing settlement-level performance is that only a proportion of pixels in a settlement actually need to be positively identified for that entire settlement to be detected. We discuss this in more detail in Section \ref{ref:section-validation}, where we describe our post-classification validation process.

\begin{figure*}[htbp]
\begin{subfigure}{\linewidth}
\centerline{\includegraphics[width=\linewidth]{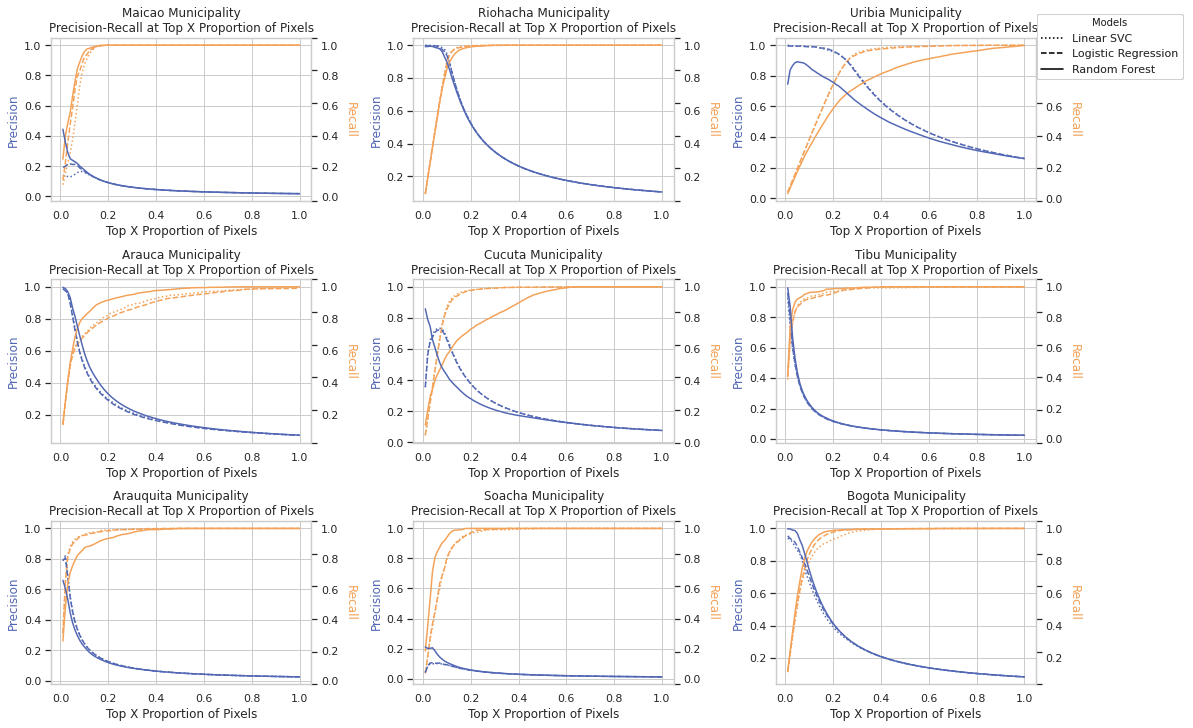}}
\caption{Pixel-level precision-recall curves at the top $x$ proportion of pixels for each of the nine municipalities.}
\label{fig:pixel-level}
\end{subfigure}

\begin{subfigure}{\linewidth}
\centerline{\includegraphics[width=\linewidth]{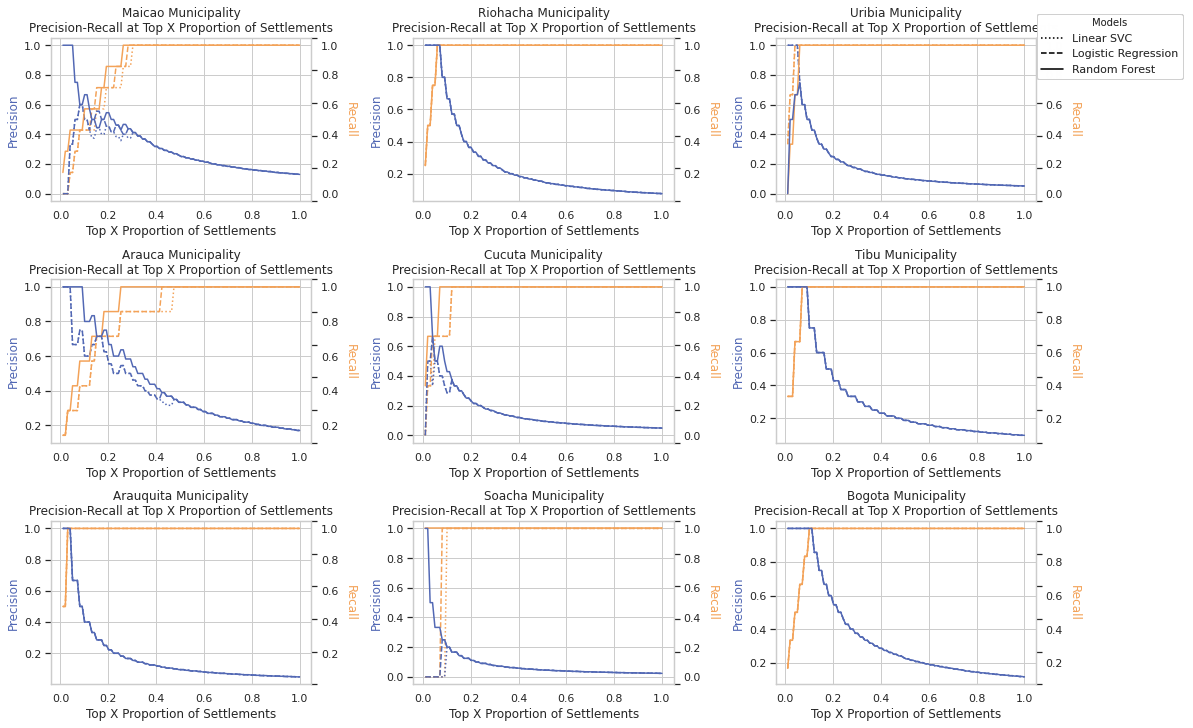}}
\caption{Settlement-level precision-recall curves at the top $x$ proportion of settlement for each of the nine municipalities.}
\label{fig:grid-level}
\end{subfigure}
\end{figure*}

For both pixel-level and settlement-level results, we compute the precision-recall curves at the top $x$ percent of pixels or settlements, where $x \in \mathbb{N}$, $x \leq 100 $. Thus, assuming that human validators begin with the highest probability pixels (or settlements) first and validate in decreasing pixel brightness, we can compute the precision and recall for any given maximum validation capacity. We present the pixel-level and settlement-level precision-recall curves for the three machine learning models in Figure \ref{fig:pixel-level} and Figure \ref{fig:grid-level} respectively.

We find that among the three models, the random forest classifier\footnote{A random forest with 800 estimators, a maximum depth of 12, a minimum of 2 samples per leaf, and a minimum of 15
samples per split using the gini criterion.} performs best in terms of both pixel-level and settlement-level performance across a majority of the nine municipalities. We also observe significantly lower precision in Maicao and Soacha compared to other municipalities as there are a large number of false positives in these two areas. In general, we identify three main sources of false positives: (1) cloud cover, (2) vegetation to bare land conversion, and (3) vegetation or bare land to formal settlement conversion. In all three cases, there is an observable change from low-intensity pixel values in earlier years, signifying vegetation cover, to high-intensity pixel values in latter years, revealing the presence of either clouds, bare land, or formal settlements. For the first case, we found that using biennial aggregates over annual aggregates helped to reduce the number of false positives due to cloud cover. For the latter two cases, we recommend using texture features in future studies as was done in \cite{wurm2017exploitation} to better differentiate informal settlement areas from formal settlements and bare land areas.

\subsection{Post-classification Validation}
\label{ref:section-validation}

In this section, we discuss our proposed two-step post-classification verification process.

\subsubsection{Remote Validation via Google Earth Pro}
Upon generating the informal settlement probability map, human validators are then tasked with manually inspecting high resolution historical satellite imagery in Google Earth Pro, starting with the brightest conglomeration of pixels, which we determine by generating 500 m x 500 m grids and calculating the mean of the top 10$\%$ of pixels per grid. Through this we are able to focus our attention on the top proportion of grids with the highest probabilities and validate in descending order of brightness. In general, we search for two main characteristics in high-resolution imagery that distinguish informal Venezuelan migrant settlements: (1) slum-like characteristics that include small roof sizes, disorganized layout of houses, and lack of nearby road structures; and (2) the absence of a settlement on date $d_1$, where $d_1$ is the earliest date for which a satellite imagery is available in Google Earth Pro and the year of $d_1$ is at least 2014, followed by the emergence of an informal settlement in that area for any date $d_2$, where $d_2 > d_1$. Note that we consider only images that are dated 2014 or later as this year marks the inception of the Venezuelan mass migration crisis. 

\begin{figure*}[h]
  \centerline{\includegraphics[width=0.7\linewidth]{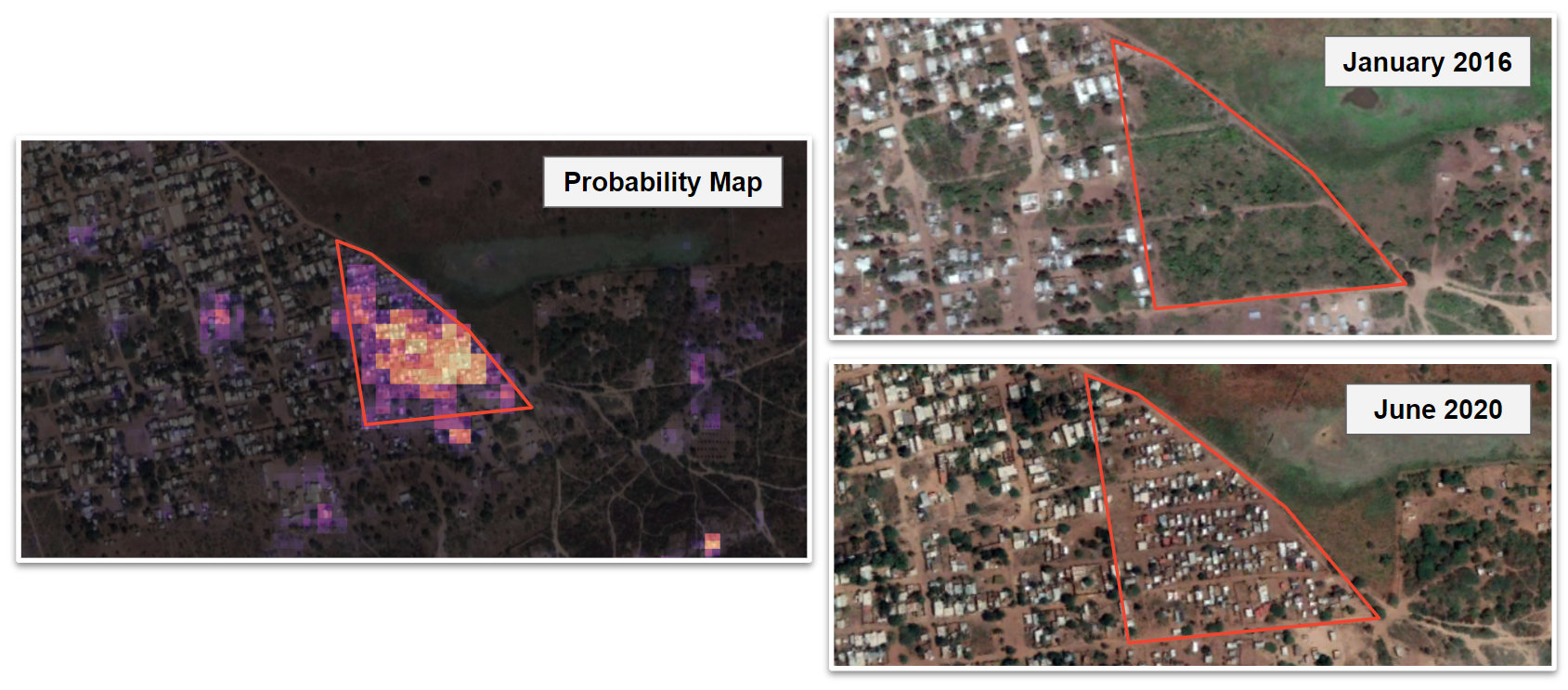}}
\caption{\textit{Remote Validation.} An example of a potential migrant settlement identified using the resulting probability map (left). Through remote inspection of historical satellite images in Google Earth Pro, we observe the emergence of a new settlement between 2016 (top right) and 2020 (bottom right)}.
\label{fig:example}
\end{figure*}

\subsubsection{On-the-Ground Validation via Premise}
\begin{figure*}[t!]
\centering
\begin{subfigure}{0.5\textwidth}
\centerline{
\includegraphics[height=1.4in]{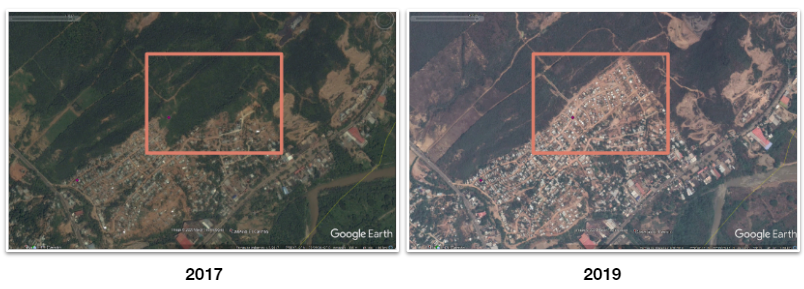}
}
\caption{}
\end{subfigure} 
\begin{subfigure}{0.5\textwidth}
\centerline{
\includegraphics[height=1.4in]{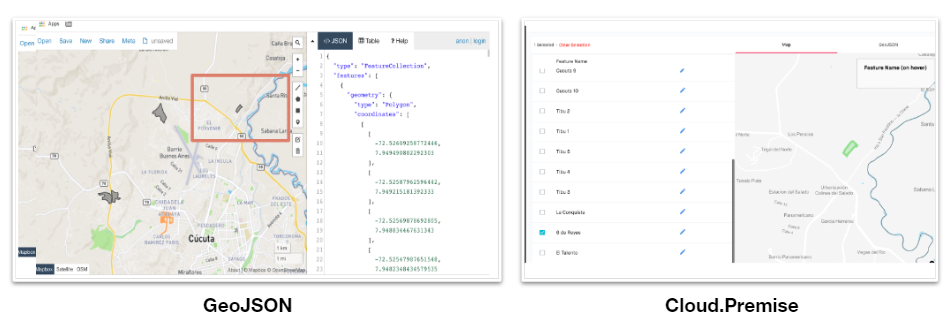}
}
\caption{}
\end{subfigure}
\par\bigskip
\begin{subfigure}[t]{0.5\textwidth}
\centerline{
\includegraphics[height=1.7in]{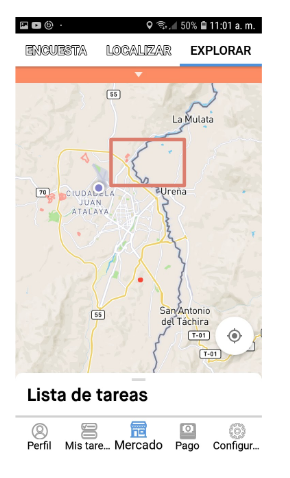}
\includegraphics[height=1.7in]{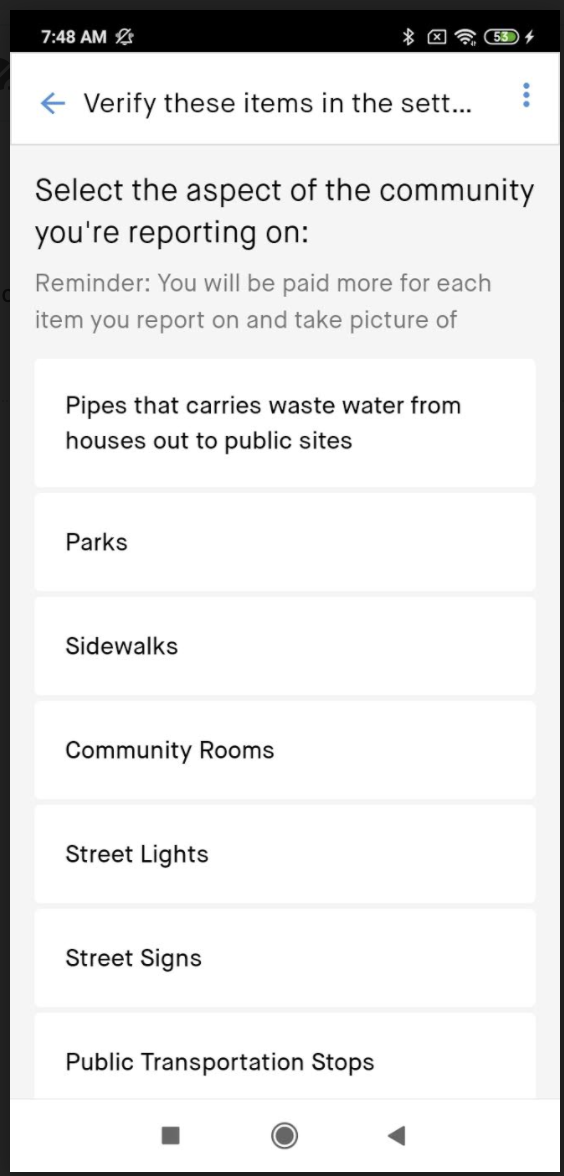}\hspace*{1mm}
\includegraphics[height=1.7in]{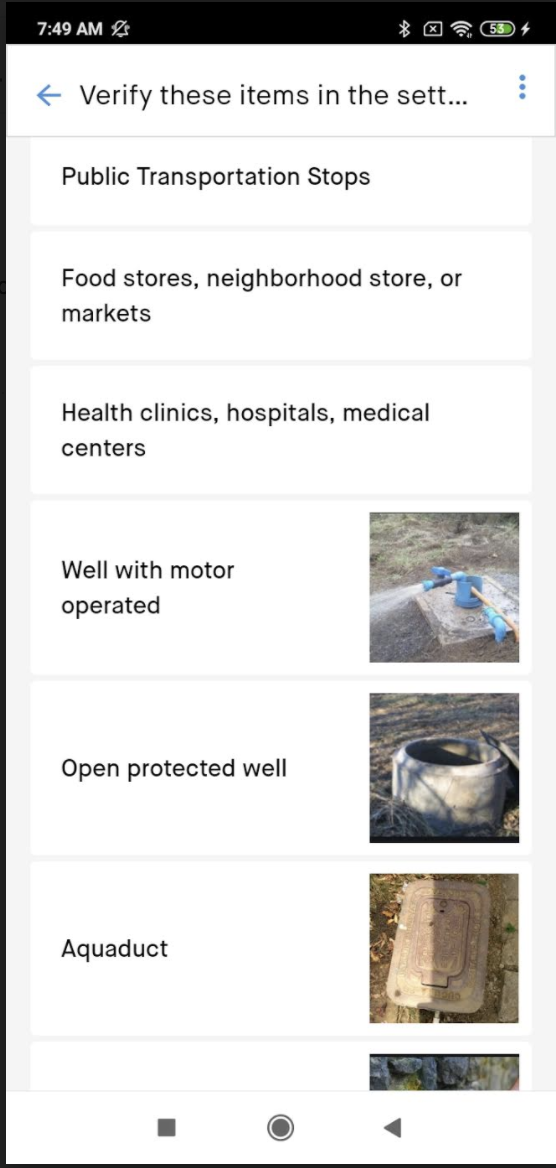}\hspace*{1mm}
\includegraphics[height=1.7in]{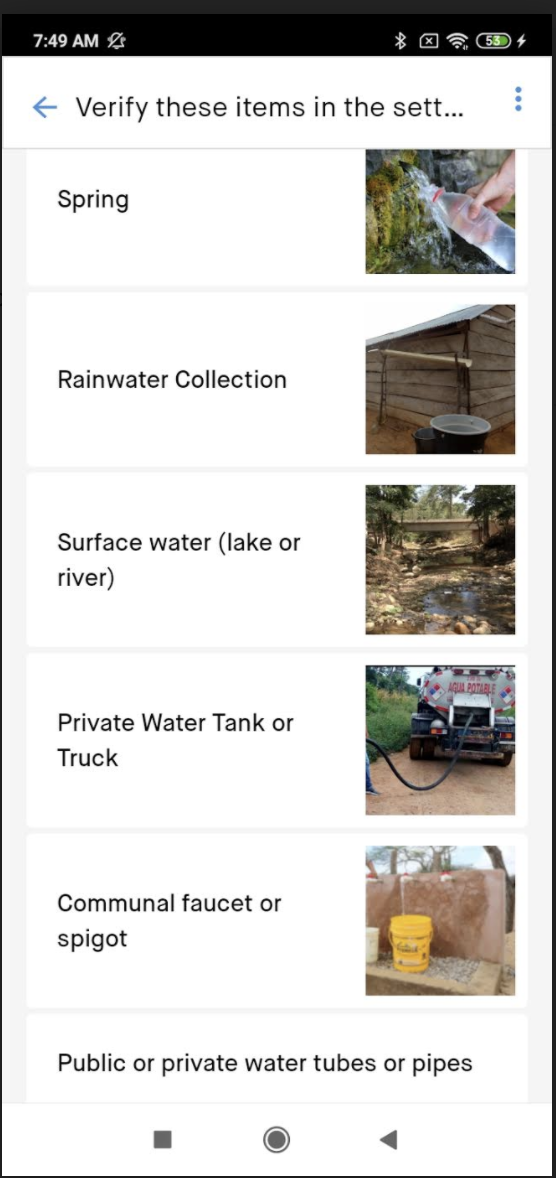}\hspace*{1mm}
\includegraphics[height=1.7in]{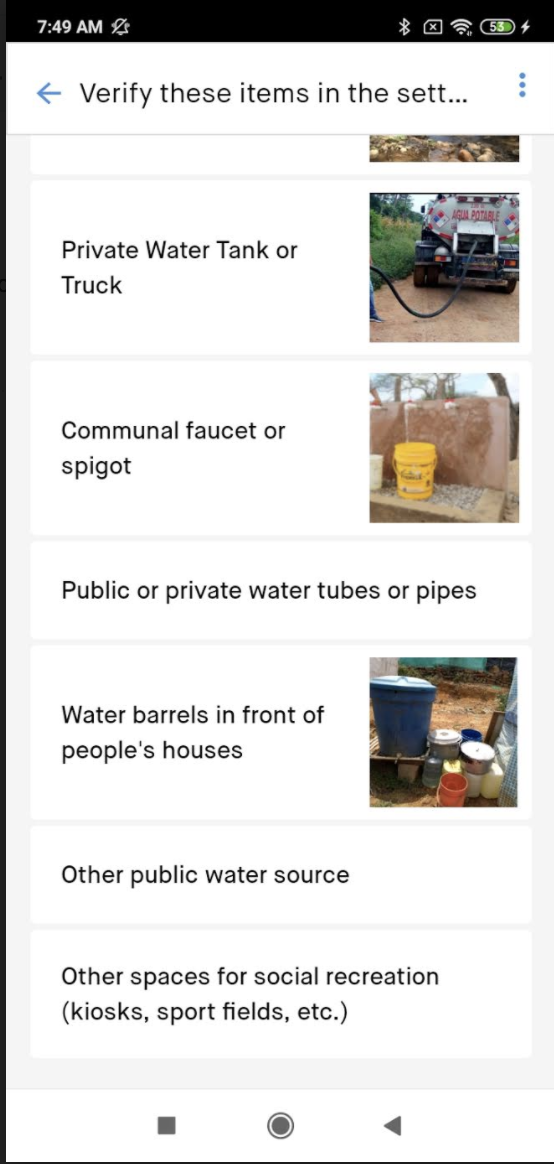}
}
\par\bigskip
\caption{}
\end{subfigure}
\begin{subfigure}{0.5\textwidth}
\centerline{
\includegraphics[height=1.7in]{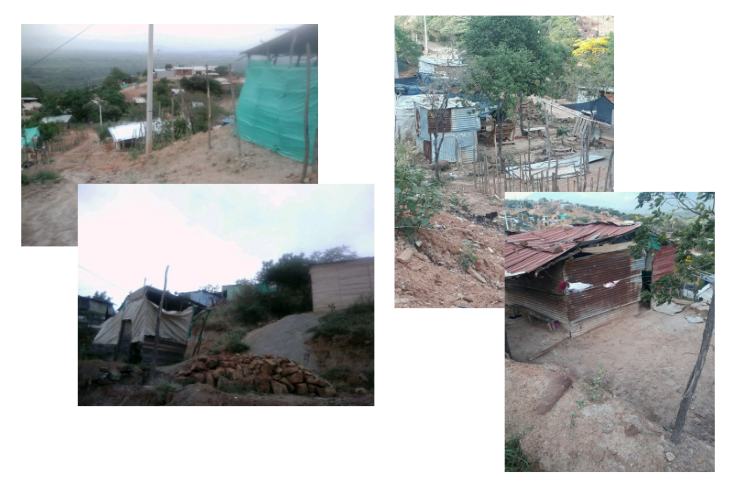}
}
\caption{}
\end{subfigure}
~
\begin{subfigure}{0.5\textwidth}
\centerline{
\includegraphics[height=1.7in]{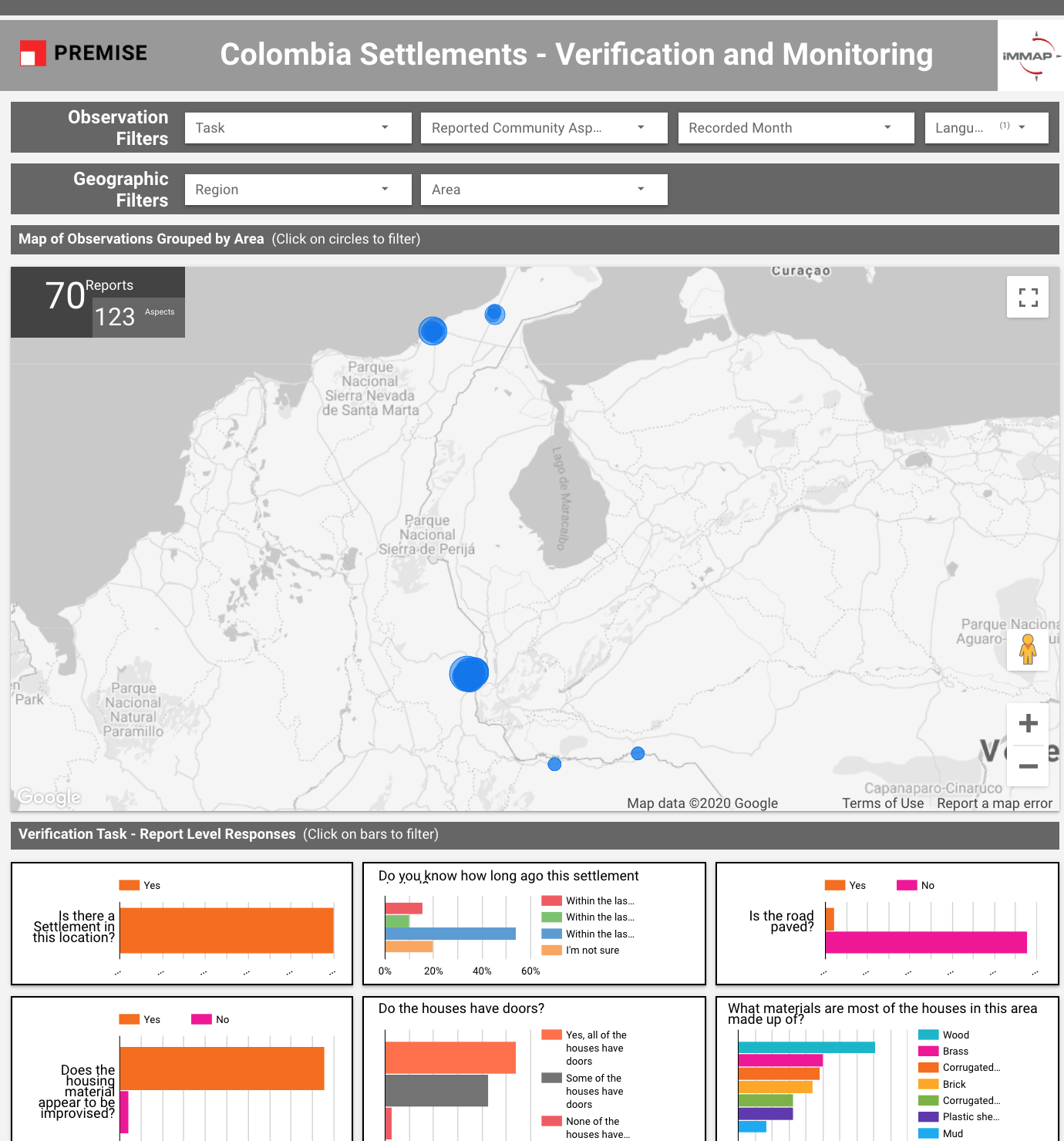}\hspace{2mm}
\includegraphics[height=1.7in]{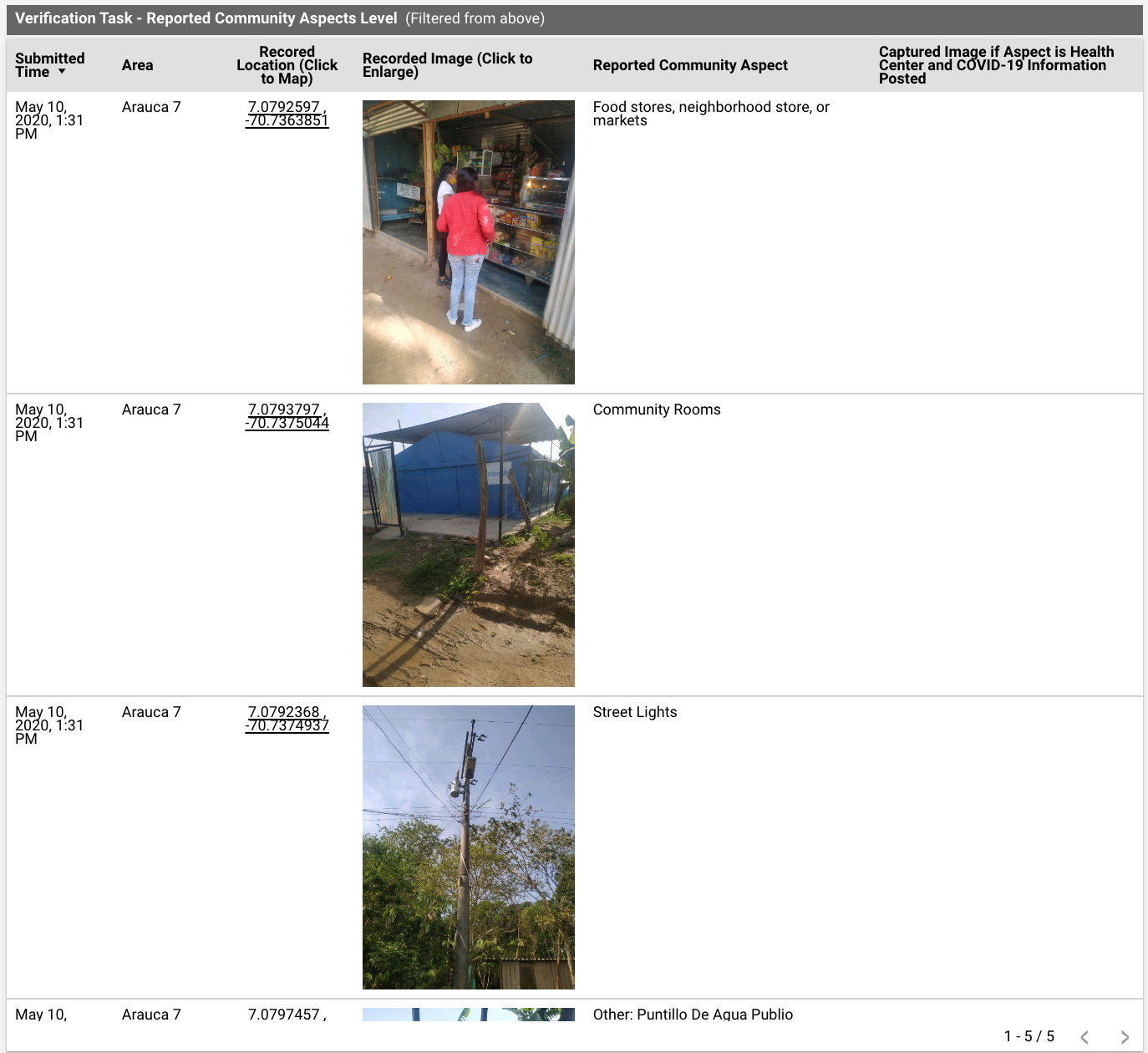}
}
\caption{}
\end{subfigure} \hspace*{1mm}

\caption{\textit{On-the-ground Validation}: Example of the validation process of a settlement in Norte de Santander through the Premise platform: (a) a settlement is identified through remote validation, (b) polygons of the settlements are drawn on GeoJSON and ingested into the Premise platform, (c) validation tasks appear in the Premise App, (d) on-the-ground Premise contributors take photos of the settlements and answer questions through the Premise App, (e) iMMAP gets access to the results through a Premise dashboard which they can use to visualize the results in aggregate and see the submitted photos by location.}
\label{fig:premise}
\end{figure*}

Once potential informal settlements are identified, we then draw vector polygons around the candidate areas using QGIS or GoogleEarth Pro. These polygons are collated and shared with our partners, Premise Data, which then enables its contributor network in the region to identify if these pre-identified settlements are actual locations where Venezuelan migrants are living. Using their proprietary app, the Premise App available on Android and iOS, Premise’s contributor network completes surveys and observations (i.e.  photographs) within these predefined polygons. The contributors are able to locate the settlements through the map shown on the mobile application and submit answers and photos that can help validate if these areas actually do house Venezuelan migrants. A second task within the app incentivizes the contributors to return to the settlements and complete a monitoring task, which focuses on identifying specific needs that the inhabitants of the settlements have with regards to water and sanitation, health, food security and overall living conditions. Figure \ref{fig:premise} demonstrates an example of the validation process for a settlement located in Norte de Santander.

%Once a potential informal settlement is identified, we then draw a vector polygon around the candidate area, using either QGIS or Google Earth Pro. These polygons are then collated into a single GeoJSON file and uploaded for use through the crowdsourcing platform Premise \cite{premise2019} for additional on-the-ground validation as shown in Figure \ref{fig:premise}. The platform helps mobilize paid data collectors to travel to the specified areas and conduct surveys and interviews to verify that the area does indeed house a community of Venezuelan migrants. Once verified, NGOs and LGUs can then provide the community with targeted humanitarian aid and assistance and continue to monitor their state of well-being over time. 

%\begin{figure}[h]
%\centerline{\includegraphics[width=1.\linewidth]{figures/2_validation.png}}
%\caption{\textit{On-the-ground Validation}: GeoJSON polygons of potential Venezuelan migrant settlements are uploaded to the Premise App, a mobile crowdsourcing platform that supports the deployment of tasks for on-the-ground validation.}
%\label{fig:premise}
%\end{figure}

\section{Conclusion}

In this study, we have tested the viability of using machine learning methods and Sentinel-2A time series satellite imagery for locating potential Venezuelan migrant settlements in Colombia that have emerged between 2015 to 2020. We modeled the problem as a supervised pixel-wise classification task and extracted a total of 66 input features per pixel consisting of raw spectral bands, derived vegetation indices, and built-up indices. Results indicate that among the models evaluated, the random forest classifier produces the best performance in terms of both the pixel-level and settlement-level precision and recall curves. Finally we have proposed a two-step verification process that includes (1) remote validation via a GIS application and (2) on-the-ground validation using the Premise App, mobile crowdsourcing platform. We are actively working with iMMAP and Premise Data to implement this validation approach in order to gather more ground truth data. The informal migrant settlements identified through this approach will be used to help LGUs, NGOs and UN agencies such as UNICEF Colombia provide targeted humanitarian aid to vulnerable Venezuelan migrant populations. The final results will also help inform planning processes such as the Humanitarian Needs Overview (HNO) 2021, by providing location-specific data around settlements of persons in need of humanitarian assistance.

\begin{acks}
This project was done as a collaborative effort between Thinking Machines, Premise Data, and iMMAP Colombia, with the financing of the Office of U.S. Foreign Disaster Assistance (OFDA) of USAID. We acknowledge the support of Pia Faustino and Ardie Orden and thank them for the insightful discussions.
\end{acks}

\bibliographystyle{ACM-Reference-Format}
\bibliography{acmart.bib}

\end{document}